\documentclass[runningheads]{llncs}
\usepackage[T1]{fontenc}
\usepackage{graphicx}
\usepackage{amsmath,amssymb,amsfonts}
\usepackage{graphicx}
\usepackage{textcomp}
\usepackage{xcolor}
\usepackage[dvipsnames]{xcolor}

\usepackage{pgf-pie}
\usepackage{algorithm}
\usepackage[noend]{algpseudocode}
\usepackage{pgfplots}
\usetikzlibrary{patterns}
\usepackage[hyphens]{url}
\usepackage[hidelinks]{hyperref}

\begin{document}
%
\title{Permuting Transactions in Ethereum Blocks: \\An Empirical Study}
%
%
\author{Jan Droll\orcidID{0009-0003-5003-3898}}
\authorrunning{J. Droll}
%
\institute{Karlsruhe Institute of Technology\\ Karlsruhe, Germany}
%
\maketitle              
\begin{abstract}
Several recent proposals implicitly or explicitly suggest making use of randomized transaction ordering within a block to mitigate centralization effects and to improve fairness in the Ethereum ecosystem.
However, transactions and blocks are subject to gas limits and protocol rules.
In a randomized transaction order, the behavior of transactions may change depending on other transactions in the same block, leading to invalid blocks and varying gas consumptions.
In this paper, we quantify and characterize protocol violations, execution errors and deviations in gas consumption of blocks and transactions to examine technical deployability.
For that, we permute and execute the transactions of over 335,000 Ethereum Mainnet blocks multiple times.
About 22\% of block permutations are invalid due to protocol violations caused by privately mined transactions or blocks close to their gas limit.
Also, almost all transactions which show execution errors under permutation but not in the original order are privately mined transactions.
Only 6\% of transactions show deviations in gas consumption and 98\% of block permutations deviate at most 10\% from their original gas consumption. 
From a technical perspective, these results suggest that randomized transaction ordering may be feasible if transaction selection is handled carefully.

\end{abstract}

\section{Introduction} 
Under Ethereum’s current block construction mechanism, block producers may select and arrange transactions at their discretion, as long as the constructed block conforms to the protocol specifications.
While the selection of transactions is designed to be driven by economic profits, the freedom to arrange transactions is nowadays used to gain additional profits, summarized under the term MEV~\cite{Daian2020FlashInstability}.
These additional profits have led to centralization in block production, facilitated censorship, and might endanger the decentralization of block validation in the long term.
A randomized transaction order might counter such centralization forces and enhance fairness in Ethereum.
In~\cite{Droll2024ShortEthereum} for example, the ordering of transactions within a block is unknown until their inclusion is guaranteed to protect users from MEV.
However, the effects of randomized transaction ordering on gas consumption and on block validity according to protocol rules remain unclear.
Other approaches, e.g.~\cite{Piet2023MEVade:Design,Bormet2024BEAT-MEV:Prevention}, propose to encrypt at least some information of transactions.
As a result, block producers would be unable to determine the gas consumption of individual transactions and the block under construction, or check blocks for protocol violations.
The concept BRAID~\cite{Resnick2024BRAID} introduces multiple entities simultaneously proposing transactions that must be included in the next block. An execution order must then be agreed upon, and dependencies between transactions may affect both gas consumption and the risk of protocol violations.
Lastly, inclusion lists (ILs), e.g.~\cite{Neuder2024UnconditionalLists}, would require block producers to include specific, e.g. censored, transactions in a block.
Again, dependencies between transactions, especially when executed at the bottom of a block as proposed in~\cite{terence2024InclusionConstraints}, may affect gas consumption and lead to protocol violations. 
Hence, we find that both the gas consumption and possible Ethereum protocol violations in unknown execution orders need investigation.

We report on the results of an experiment that permutes and executes transactions of Ethereum Mainnet blocks to analyze protocol violations and the gas consumption of blocks and transactions. 
In particular, we answer the following questions with an empirical study of more than 335,000 Ethereum Mainnet blocks: 

\begin{itemize}
    \item How many block permutations violate current protocol rules, rendering the block permutation invalid, and for what reason?
    \item How many blocks and transactions deviate from their original gas consumption when permuted?
    \item How large is the gas deviation of blocks and transactions when permuted?
\end{itemize}

Permuting transactions in blocks is close to an unknown execution order and activates dependencies between transactions.
The presented approach to permute transactions in Ethereum blocks avoids protocol violation at the cost of only minor, backward-compatible modifications of block validation rules.
We also analyze execution errors that, unlike protocol violations, affect only individual transactions and user experience.

The remainder of the paper is structured as follows:
Section~\ref{ethereum} briefly introduces concepts of Ethereum required to understand this paper and differentiates this work from existing work before presenting our methodology in Section~\ref{experiment}.
After presenting the result of the experiment in Section~\ref{results}, we discuss the findings in Section~\ref{discussion} and finally draw our conclusion in Section~\ref{conclusion}.

\section{Fundamentals and Related Work}
\label{ethereum}
The Ethereum~\cite{Buterin2014APlatform} network operates a replicated state machine everyone is able to interact with.
Interactions are expressed in form of transactions, i.e., signed statements that transition the state of the replicated state machine during their execution.
Transactions are executed in the order they are arranged in a block. 
Executing a block is equivalent to consecutively executing all contained transactions on the state that results from processing previous blocks in the blockchain.

State transitions depend on a transactions' data and follow deterministic rules specified as instructions.
Instructions are in particular capable to read from and write to the state, support logical loops, and conditional execution branching.

For each executed instruction, a specific amount of gas, an abstract unit for computation and storage costs, is due.
Transactions specify a limit for the cumulative amount of gas that can be consumed during their execution together with a conversion rate between gas and Ether, Ethereum's native currency. 
Based on the actually consumed gas $g$ and specified conversion rate $p$, the transaction fee $f = g  \cdot p$ is deducted from the account of the transaction sender.

Accounts are objects, identified by addresses, that define the state of the replicated state machine and are used to keep track of individual Ether balances in Ethereum.
In addition, accounts record the number of executed transactions with the account's address specified as sender up to the current state, referred to as \textit{nonce}.
Transactions also specify a number called \textit{nonce} that must match the nonce recorded in the transaction sender's account on execution.
So-called smart contract accounts contain instructions organized into functions, which are executed when triggered by transactions.

A block violates the Ethereum protocol rules if during the block's execution the nonce of a transaction does not equal the nonce of the senders account, the transaction sender owns less Ether than required to settle maximum transaction fees plus the amount of ether to be transferred, or if the cumulative gas used by transactions exceeds the block gas limit.
Note that the maximum transaction fee $f^\mathrm{max} = l \cdot p$ is calculated from the gas price $p$ and gas limit $l$ as specified by the transaction instead of the actually consumed gas $g$.

The gas consumption of Ethereum transactions was investigated in~\cite{Zarir2021} with the goal to determine the stability of the gas consumption of smart contract functions.
The authors analyzed the on-chain recorded execution results of transactions between October 2017 and February 2019 and found that 25\% of functions are unstable, i.e., show significantly varying gas consumptions among invocations by transactions.
In addition, 50\% of all transactions invoke one of these unstable functions.
We analyze the gas consumption along with execution and protocol errors primarily based on permuted execution results, not on on-chain executions.

A related field of research is to estimate the actual gas consumption of single individual transactions.
This estimation is required during transaction creation to delimit the amount of spendable gas while avoiding `out-of-gas' errors during transaction execution.
Common strategies involve compiler outputs, e.g. as provided by solc\footnote{https://docs.soliditylang.org/en/latest/using-the-compiler.html {\scriptsize (2025-05-20)}}, and binary searches, e.g. when calling `eth\_estimateGas' on go-ethereum\footnote{https://github.com/ethereum/go-ethereum {\scriptsize (2025-05-20)}}.
The authors of \cite{Soto2020FuzzingContracts,Li2021GasEthereum} deal with optimizing such estimations.
However, our work does not focus on estimating gas for new transactions, instead we analyze if limits are well chosen.

Permutations have been used for analysis purposes in~\cite{Adams2024TheExchanges,Angeris2023TheValue}.
The papers aim to quantify MEV by comparing the cost for users across block permutations. 
While~\cite{Angeris2023TheValue} is limited to a formal definition based on permutations,~\cite{Adams2024TheExchanges} also conduct an experiment.
They analyzed up to 16 permutations, but only of specific sets of MEV related transactions, to estimate a value for their `reordering slippage' metric.
In contrast, we quantify the technical sensitivity of Ethereum regarding protocol violations, execution errors and gas consumption when entire blocks are permuted - based on thousand permutations per block.


Some non-Ethereum systems, e.g. Solana~\cite{YakovenkoSolana:Reader}, optimize transaction ordering for parallel execution, taking state-dependencies into account.
Rather than optimizing transaction ordering, we empirically analyze the effects of state and ordering dependencies in Ethereum.

\section{Methodology}
\label{experiment}
We conducted an experiment to analyze the effects of permuted transaction execution on block validity and on the gas consumption of both blocks and transactions.
In the experiment, the transactions of Ethereum Mainnet blocks are permuted and the results of the executions are recorded for analysis. 

\subsection{Setup and Workflow}
\label{workflow}
\begin{figure}[htbp]
    \centering
    \includegraphics[width=7cm]{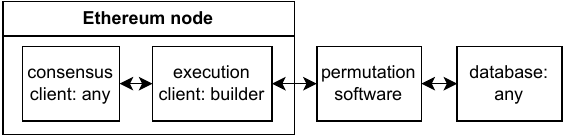}
    \caption{Components required for the experiment.
    Arrows represent communication between components.
    The execution client `builder' and our own permutation software are the experiment's core components.
    Any standard consensus client is suitable for the experiment and the permutation software is database agnostic.
    }
    \label{figComponents}
\end{figure}

The technical setup, illustrated in Fig.~\ref{figComponents}, is made up of an Ethereum node, consisting of a standard consensus client and a modified execution client, a permutation software, and a database.
The consensus client ensures that the execution client follows the canonical blockchain of the Ethereum Mainnet. 
Regular executions clients are responsible to maintain the state of the replicated state machine based on the canonical blockchain.
The modified execution client used in our experiment provides additional API functionality to execute custom blocks.
Custom blocks consists of a list of transactions together with a state number\footnote{The state with number $n$ is the resulting state after processing the block with number $n$ in the chain.} and other essential information.
Transactions in the custom block are consecutively executed on the specified state in the given order, but the performed state transitions are ephemeral for the time of the execution of the custom block.
The consecutive execution is crucial for our experiment to guarantee interdependent execution of transactions. 
However, for the experiment, we used only functionality of the standardized `eth' API provided by regular execution clients, except for the execution of custom blocks.
The permutation software creates custom blocks by permuting transactions of blocks received from the Ethereum node, instructs the modified execution client to execute these custom blocks, and records the results in a database used for further analysis.

For the experiment, we used `builder'\footnote{https://github.com/flashbots/builder {\scriptsize (2025-05-20)}}, a modified version of `go-ethereum'\footnote{https://github.com/ethereum/go-ethereum {\scriptsize (2025-05-20)}}, as execution client. 
The permutation software is implemented by ourself and public available~\cite{Droll2025TxPerm}. 

The workflow of the permutation software (in the following simply called `software') is as follows:
After successfully connecting to the Ethereum node, the software subscribes to new block heads.
On receiving a new head, the block processing starts by requesting the full block referred to in the header together with receipts of each transaction included in the block from the Ethereum node.

\paragraph{Re-execution}
Then, the software requests the execution client to re-execute the block as a custom block.
This custom block is identical to the received full block, which means that it consists of all transactions of the previously received full block in the identical order.
The purpose of the re-execution is to verify the correct behavior of the execution client with the modifications in place.
The execution client either responds with a so called \textit{core error} or the result of the re-execution.
A core error means that the block violates a rule specified in the Ethereum specification and the custom block execution aborted because the block would be invalid.
In case of an core error, the software extracts the concrete type of the core error and during which transaction's execution the error occurred, and records both the type and the transaction in the database. 
In addition, any further processing of the current block is aborted.
Otherwise, the software extracts the amount of gas consumed by each transaction during the re-execution together with possible execution errors from the result and stores both with the gas consumed in the original block, extracted from the receipts, in the database.
In addition, some information regarding the block, including the gas consumed as declared in the header is stored in the database.

\paragraph{Permuted execution}
Now, depending on the content of the block, the software either randomly permutes all transactions in the block with the algorithm described in Sec.~\ref{permutation} to create custom blocks or the software computes a custom block for each possible permutation.
In either case, the software requests the execution client to execute the now permuted block as custom block.
In case of an error response, the type of the core error and the transaction that caused the core error is extracted and recorded in the database.
Further processing of the block continues, which is a different error handling compared to the re-execution in which block processing aborts.
Otherwise, the amount of gas consumed by each transaction is recorded in the database and, if the execution of a transaction resulted in an so called \textit{execution error}, the type of the execution error is also extracted and recoded in the database.
Execution errors do not result in invalid blocks, which is the difference to core errors. 
We randomly permute blocks 1000 times except for blocks that contain less than seven distinct transaction senders.

\begin{figure}[htb]
    \centering
    \includegraphics[width=10cm]{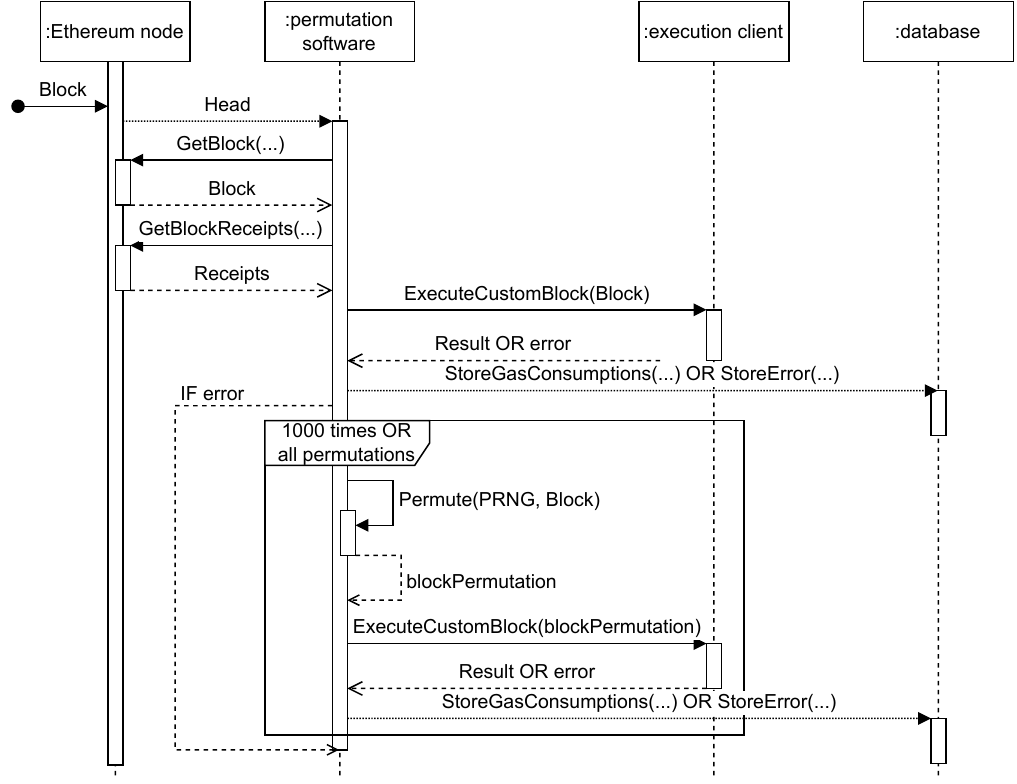}
    \caption{Block processing workflow.
    Interaction between permutation software and Ethereum node use the standardized `eth' API.
    Interaction between the permutation software and the execution client utilize the additional custom block functionality.
    In practice, the execution client is part of the Ethereum node as depicted in Fig.~\ref{figComponents}.
    }
    \label{figWorkflow}
\end{figure}

In pre-production runs, we discovered two incorrect behaviors of builder and corrected them before conducting the experiment:
First, for 0.2\% of analyzed transactions (13\% of blocks) the state transitions was incorrect because the API did not provide an option to set the header field `mixHash' in the header, even transactions can access the field's value during execution. See EIP-4399~\cite{Kalinin2021EIP-4399:PREVRANDAO} for more details.
Second, the builder did not implement the EIP-4788~\cite{Stokes2022EIP-4788:EVM} oracle that provides the root of the parent beacon block to the EVM.
Some transactions, likely all related to EigenLayer, show a different gas consumption if the oracle is not implemented.

\subsection{Permutation}
\label{permutation}
We designed the permutation algorithm to keep the relative order of transactions by a single sender unchanged.
This design avoids protocol violations (see Sec.~\ref{ethereum}) due to a mismatch of the nonce specified by the transaction and in the current state\footnote{A transaction execution increases the nonce within the account.}, and the implementation is straightforward.

The resulting algorithm is provided in Algorithm~\ref{permutationAlgorithm} and works as follows:
First, the sender of each transaction is recovered from the transactions' signature in the order of the original block.
The sender address is appended to a list if not already present to avoid duplicates.
Then, the list of addresses is shuffled using the Fisher-Yates shuffle algorithm as implemented in the Golang standard library `math/rand'\cite{GoogleShuffleCode}.
If the block contains less than seven distinct senders, the algorithm deterministically computes each possible permutation of the list of addresses (which are at most 720 permutations) instead of shuffling the list.

\begin{algorithm}
\caption{Permutation algorithm}\label{euclid}
\label{permutationAlgorithm}
\begin{algorithmic}[1]
\small
\Procedure{PermuteTransactions}{}
\State input: \textit{txs} // list of transactions
\State input: \textit{seed} // number

\State var $\textit{distinctSenders}$ // emtpy list
\For{$i \gets 0; i < len(txs); i++$}
\State $\textit{sender} \gets \textit{ecrecover(txs[i])}$ // Ethereum precompile
\If {$\textit{sender} \notin \textit{distinctSenders}$} 
\State $ \textit{append}(\textit{distinctSenders}, \textit{sender})$
\EndIf
\EndFor
\State $\textit{permutedSenders} \gets \textit{shuffle}(\textit{seed}, \textit{distinctSenders})$ // Fischer-Yates

\State var $\textit{permutedTxs}$ // emtpy list
\For{$j \gets 0; j < len(\textit{permutedSenders}); j++$}
\State $\textit{permutedSender} \gets \textit{permutedSenders}[j]$
\For{$k \gets 0; k < len(txs); k++$}
\State $\textit{originalSender} \gets \textit{ecrecover(txs[k])}$
\If {$\textit{permutedSender} = \textit{originalSender}$}
\State $ \textit{append}(permutedTxs, txs[k])$
\EndIf
\EndFor
\EndFor

\State \textbf{return} \emph{permutedTxs}.
\EndProcedure
\end{algorithmic}
\end{algorithm}

The Fischer-Yates algorithm guarantees that every permutation of the list is equally likely~\cite{KnuthDonaldErvin1998TheProgramming.}.
The final list of permuted transactions is then incrementally constructed by iterating over the now shuffled list of distinct sender addresses and thereby appending each transaction of the sender from the original list of transactions. 

The Fisher-Yates algorithm requires a source of randomness to randomly shuffle the list of transactions.
In the experiment we used seed-based pseudo-random number generators (PRNG) so that the experiment is reproducible.
We instantiated an independent PRNG for each block (block-level PRNG) with the hash of the respective block as seed.
This block-level PRNG is then used to seed an independent PRNG for each permutation (permutation-specific PRNG) with a freshly drawn pseudo-random value.
The permutation-specific PRNG is then used as source of randomness by the Fisher-Yates algorithm.

\subsection{Ethical considerations}
The experiment uses only public data, namely some blocks of the Ethereum Mainnet.
This public data consists of pseudonymous data in form of cryptographic addresses, and further, potentially personal, data supplied by transactions senders or block builders.
The experiment's database contains only data that is stored and processed by every Ethereum node together with permutation results.
Analyzing the permutation results involves only data processed and stored by every Ethereum node and a list of transactions from analysis by~\cite{Guru2025}.
No attempts are made to deanonymize data and no additional new data links are created beyond the one to~\cite{Guru2025}.
Therefore, we do not see ethical aspects that originate from performing the experiment.

\section{Results}
\label{results}
With the experiment described in the previous section, we collected data for about six weeks during March and April 2025 (blocks 22039048 to 22374561).
During that period, we observed 335,886 blocks\footnote{We observed not only canonical blocks, but also blocks that got not finalized.} from which 335,646 blocks contained transactions.
We permuted only the latter blocks, so we refer to them as `permuted blocks', and those blocks contained more than 56 million transactions.
About 64\% of transactions interact with smart contracts while the remaining transactions are simple Ether transfers.
In 1.8 million cases, the same smart contract was invoked by at least two transactions within a single block. 
However, we expect even higher interdependencies, as smart contracts can directly interact with one another.
In this section, with `original', as in `original block`, we refer to data exactly as agreed on by the Ethereum network and persisted in the blockchain.
In the experiment, we recorded the gas consumptions for more than 42 billion permuted transaction executions, and for more than 72 million block permutations, we recorded the reasons for the their invalidity.
Note that a permuted block has typically up to 1000 block permutations from which some executed successfully and some were invalid due to protocol violations.
For 628 blocks we computed all possible permutations, because they contained less than 7 distinct transaction senders.
In the following, we characterize blocks and transactions based on the results of the experiment, starting with protocol violations (Sec.~\ref{protocolViolations}), then analyze execution errors (Sec.~\ref{executionErrors}) and finally covering the gas consumption of blocks and transactions (Sec.\ref{behavior}).

\begin{figure}[htbp]
    \centering
    \begin{tikzpicture}
\begin{axis}[
    height=5cm,
    width=8.7cm,
    log origin=infty,
    ytick={0, 0.0001, 0.001, 0.01, 0.1, 1.0, 10.0, 100.0},   
    ylabel={Number of blocks},                  
    ylabel style={yshift=5pt},       
    xlabel={Fraction of invalid permutations per block},                 
    xlabel style={yshift=-50pt, xshift=-5pt},       
    xtick={0, 1, 2, 3, 4, 5, 6, 7, 8, 9, 10, 11, 12, 13, 14, 15, 16, 17, 18, 19, 20, 21},              
    xticklabels={
    \text{0\% (never)},
    \text{(0\%, 10\%]},
    \text{(10\%, 20\%]},
    \text{(20\%, 30\%]},
    \text{(30\%, 40\%]},
    \text{(40\%, 50\%]},
    \text{(50\%, 60\%]},
    \text{(60\%, 70\%]},
    \text{(70\%, 80\%]},
    \text{(80\%, 90\%]}, 
    \text{(90\%, 100\%)},
    \text{100\% (always)}
    }, 
    yticklabels={0.0001\%, 0.001\%, 0.01\%, 0.1\%, 1\%, 10\%, 100\%}, 
    ymin=0, ymax=100,                  
    bar width=6pt,                    
    ymajorgrids=true,                  
    grid style=dashed,                  
    xticklabel style={rotate=90},      
    ybar,                               
    ymode=log,
    nodes near coords,                 
    nodes near coords align={vertical},
    every node near coord/.append style={font=\tiny, rotate=90, xshift=17px, yshift=-5px}, 
    point meta=rawy,
    legend style={at={(0.63,0.19)}, anchor=north, legend columns=-1}
]
\addplot coordinates { 
    (0, 65.12) 
    (1, 1.73) 
    (2, 1.31) 
    (3, 1.09) 
    (4, 0.93) 
    (5, 6.9) 
    (6, 7.22) 
    (7, 1.45) 
    (8, 3.25) 
    (9, 2.65) 
    (10, 8.14) 
    (11, 0.21) 
};
\end{axis}
\end{tikzpicture}
\caption{
Distribution of blocks according to the fraction of core errors observed in their permutations. 
Note the log-scaled y-axis and x-axis interval ranges.
}
    \label{figDistributionInvalidPermutationsAmongPermutedBlocks}
\end{figure}
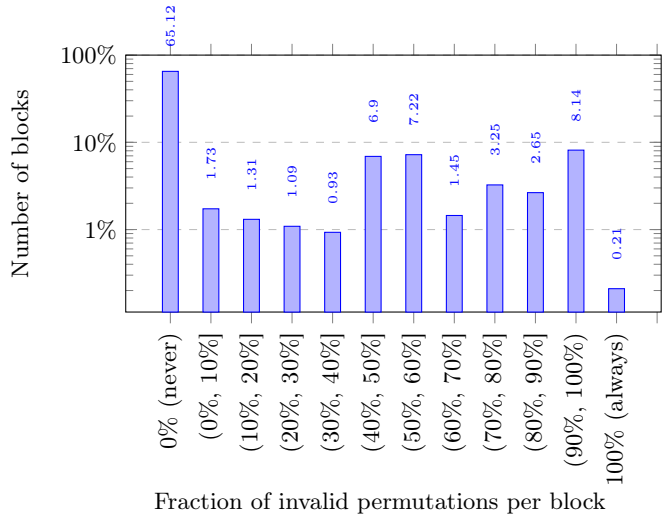

\subsection{Protocol Violations}
\label{protocolViolations}
While we designed the permutation algorithm for the experiment in a way that eliminates not matching nonces in block permutations, we observed two other protocol violations in form of core errors which invalidated 22\% of all block permutations, affecting 35\% of blocks. 
In Fig.~\ref{figDistributionInvalidPermutationsAmongPermutedBlocks}, the fraction of invalid block permutations per permuted block is shown as a distribution across all permuted blocks.
For 65\% of permuted blocks, all block permutations were valid while for 0.21\% of permuted blocks, all block permutations were invalid.
With 71\%, most invalid block permutations observe an \textit{insufficient funds} error which can be related to privately mined transactions.
The remaining 29\% invalid block permutations run into an \textit{gas limit reached} error that could be avoided by applying a gas margin.
Only 4.6\% of blocks have both gas limit reached and insufficient funds errors in permutations. 
Invalid block permutations could be avoided by restraining to mine transactions privately and respecting a gas margin.


Insufficient funds errors occur if the sending account of a transaction owns not enough Ether to settle the maximum transaction fee $f^\mathrm{max}$ together with the amount of Ether to be transferred.
Two scenarios can cause this kind of core error in block permutations: Either the sending account received some Ether\footnote{From itself, another external owned account or a smart contract.} by an earlier transaction in the original block or a transaction $\alpha$ caused a higher gas consumption of a transaction $\beta$ sent by account $\sigma$ so that $\sigma$ has no more sufficient funds for a subsequent transaction $\gamma$.
Future work can analyze the frequency of such scenarios by analyzing intermediate states of executions.
However, 99.93\% of transactions with insufficient funds in permutations can be attributed to privately mined transactions according to data from~\cite{Guru2025}.
Privately mined transactions are first seen in a block and not on the network.

Gas limit reached errors occur if the execution of a transaction exceeds the gas limit of the block.
Section~\ref{gasDeviation} demonstrates that some transactions require more gas in certain block permutations than in the original block, thereby potentially causing this error.
There is a moderate to strong positive correlation (0.58) between the extend to which the block gas limit is used in the original block and the ratio of gas limit reached errors in permutations of a block.
For example, 70\% of the permuted blocks with gas limit reached errors consumed at least 91\% of their gas limit in the original block.
Also, less than 0.3\% of permuted blocks with gas limit reached errors that utilize at most half of the gas limit (The so called \textit{target} since EIP-1559) in the original block observe an invalidation due to reaching the block gas limit in our experiment.
According to these findings, the closer blocks are to their block gas limit, the more likely they will be invalid when permuted due to gas limit reached core errors. 

\subsection{Execution Errors} 
\label{executionErrors}
We tracked execution errors\footnote{\url{https://github.com/ethereum/execution-specs/blob/master/src/ethereum/cancun/vm/exceptions.py} \scriptsize{(2025-05-20)}} that affect only the transaction causing the error, and not the validity of block permutations.
We expected more failing transactions due to state dependencies and found that those contributing to the increased failure rate tend to send Ether directly to the block assembler and are privately mined.
Also supported by the observation that these transactions fail due to a revert instruction, we conjecture they are MEV-related, and believe regular users' experience is not negatively affected by random transaction ordering. 

Execution errors occur in 81\% of original blocks, increasing to 94\% under permutation.
This trend is even more prevalent in transactions where 2.7\% of transactions fail due to an execution error in original blocks and 4.7\% in block permutations.
The increase is reflected by new failing transactions (transactions which do not fail in the original block, but in a block permutation), only a small number of failing transactions in the original block remain non-failing across all permutations.
One might argue that with a randomized transaction order, execution errors occur more frequently due to new failing transactions, while transactions that failed in the original block will continue to fail in block permutations.
Such a statement is only partially supported because the latter transactions fail in 72\% of their executions in block permutations on average and new failing transactions in 42\% on average.

More interestingly, all new failing transactions share a common property.
They either directly affect the coinbase address of the block in a positive way (send Ether directly to the block producer) or they do not pay fees beyond the base fee.\footnote{Fees beyond the base fees might be ignored as more than 99.99\% of new failing transactions share the first property.}
Only 2\% of transactions with this property are new failing transactions.
While not paying fees beyond the base fee is unremarkable, transferring Ether to the block producer draws attention.
The effect of a direct payment to block producer is that the sender of the transactions can decide that only specific block producers profit from their transaction.
We found that 99\% of the transactions that send Ether directly to the block producer are privately mined according to~\cite{Guru2025}. 
Several institutions offer services that result in privately mined transactions, for example to protect from MEV extraction but also to protect MEV extraction itself, and such services guarantee a specific order of transactions.
We conjecture that new failing transactions are related to so called MEV extraction bundles without providing further evidence or elaborating details. 

In both original blocks and block permutations, we observed seven different execution error types.
74\% of execution errors in block permutations (66\% in original blocks) are revert errors, meaning that the logic of the application invoked by a transaction executed a special instruction that causes all state modifications made by applications to be rolled back.
In 23\% of execution errors in block permutations (31\% in original blocks), the gas limit of the transaction was reached.
In 2.5\% of execution errors in block permutations (3.4\% in original blocks), an invalid instruction was attempted to be executed.
The remaining 0.04\% of execution errors in block permutations (0.03\% in original blocks) correspond to scenarios in which the stack of the EVM underflowed, a variable to track the gas consumption overflowed, the maximum code size was exceeded or an application tried to jump to an invalid destination.
We highlight that the execution error type of new failing transactions is a revert error in 99.7\% of execution errors and the gas limit of such transactions was reached in 0.02\%.

\subsection{Gas Deviations}
\label{gasDeviation}
The purpose of the experiment described in Sec.~\ref{experiment} was also to quantify the variations in gas consumption of blocks and transactions. 
We break down the results by blocks and transactions at first and then across all recorded executions, without associating them to individual blocks and transactions. 



\subsubsection{Block and Transaction Associated Results}
\label{behavior}
In this section, we evaluate data per block and per transaction:
We calculate our metrics for each block and transaction with the executions of the respective block or transaction.
The visualizations, however, plot the metric for each individual blocks and transactions.

\paragraph{Fraction of Deviations}
The first metric $f^\mathrm{D}$ measures the fraction of executions with gas deviations relative to the total number of successful executions $N$ of a block $B$ or transaction $T$ according expressions~\eqref{eqFracDivBlockPerms} and~\eqref{eqFracDivTxExecs}.
Note that successful executions are block permutations without protocol violations.

\begin{minipage}{0.45\linewidth}
\begin{equation}
\label{eqFracDivBlockPerms}
    f^\mathrm{D}_B = \frac{n^\mathrm{D}_B}{N_B}
\end{equation}
\end{minipage}
\begin{minipage}{0.45\linewidth}
\begin{equation}
\label{eqFracDivTxExecs}
    f^\mathrm{D}_T = \frac{n^\mathrm{D}_T}{N_T}
\end{equation}
\end{minipage}

\begin{figure}[htbp]
    \centering
    \begin{tikzpicture}
\begin{axis}[
    height=5cm, 
    width=8.7cm, 
    log origin=infty,
    ytick={0, 0.0001, 0.001, 0.01, 0.1, 1.0, 10.0, 100.0},   
    ylabel={Share of transactions/blocks},                  
    ylabel style={yshift=10pt},       
    xlabel={Fraction of deviating permutations $f^\mathrm{D}_T$ / $f^\mathrm{D}_B$},                 
    xlabel style={yshift=-45pt, xshift=-15pt},       
    xtick={0, 1, 2, 3, 4, 5, 6, 7, 8, 9, 10, 11, 12, 13, 14, 15, 16, 17, 18, 19, 20, 21},              
    xticklabels={
    \text{0\% (never)},
    \text{(0\%, 10\%]},
    \text{(10\%, 20\%]},
    \text{(20\%, 30\%]},
    \text{(30\%, 40\%]},
    \text{(40\%, 50\%]},
    \text{(50\%, 60\%]},
    \text{(60\%, 70\%]},
    \text{(70\%, 80\%]},
    \text{(80\%, 90\%]}, 
    \text{(90\%, 100\%)},
    \text{100\% (always)}
    }, 
    yticklabels={0.0001\%, 0.001\%, 0.01\%, 0.1\%, 1\%, 10\%, 100\%}, 
    ymin=0, ymax=100,                  
    bar width=6pt,                    
    ymajorgrids=true,                  
    grid style=dashed,                  
    xticklabel style={rotate=90},      
    ybar,                               
    ymode=log,
    nodes near coords,                 
    nodes near coords align={vertical},
    every node near coord/.append style={font=\tiny, rotate=90, xshift=17px, yshift=-5px
    }, 
    point meta=rawy,
    legend style={at={(0.63,0.19)}, anchor=north, legend columns=-1}
]
\addplot coordinates { 
    (0, 11.62) 
    (1, 0.05) 
    (2, 0.06) 
    (3, 0.06) 
    (4, 0.24) 
    (5, 4.02) 
    (6, 3.97) 
    (7, 1.48) 
    (8, 5.28) 
    (9, 6.33) 
    (10, 21.65) 
    (11, 45.23) 
};
\addplot coordinates { 
    (0, 94.0) 
    (1, 0.11) 
    (2, 0.25) 
    (3, 0.23) 
    (4, 0.34) 
    (5, 1.44) 
    (6, 1.42) 
    (7, 0.58) 
    (8, 0.33) 
    (9, 0.48) 
    (10, 0.44) 
    (11, 0.38) 
};
\legend{Blocks, Transactions}
\end{axis}
\end{tikzpicture}
\caption{Distribution of blocks (blue) and transactions (red) according to the fraction of deviating block / transaction permutations ($f^\mathrm{D}_B$ or $f^\mathrm{D}_T$) observed in their permutations. Note the log-scaled y-axis and x-axis interval ranges.
}
    \label{figShareTransactionExecutionsWithDeviations}
\end{figure}
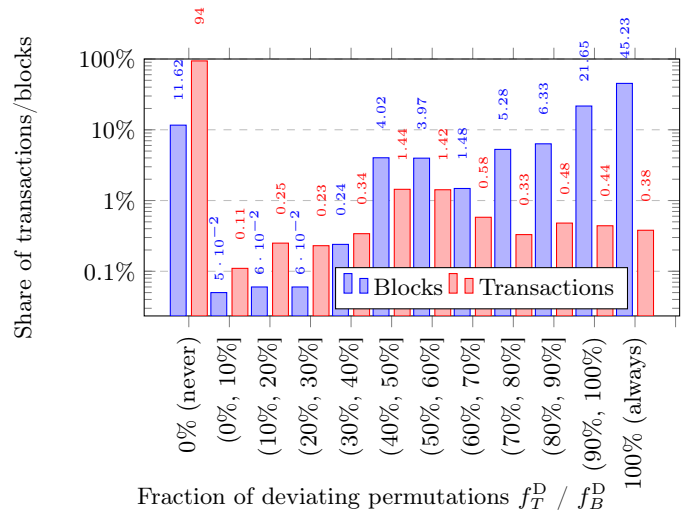

Values for both $f^\mathrm{D}_B$ and $f^\mathrm{D}_T$ must be in the interval $[0, 1]$ since between zero and $N$ executions can deviate.
The metric is calculated for each permuted block and transaction in our database and Fig.~\ref{figShareTransactionExecutionsWithDeviations} plots $f^\mathrm{D}_B$ for all 335,646 permuted blocks as well as 
$f^\mathrm{D}_T$ for all 56 million transactions.

Over 45\% of permuted blocks deviate in all permutations from their original gas consumption in our experiment, while about 11\% of permuted blocks never deviate.
However, permuted blocks clearly tend to deviate from their original gas consumption.
In contrast, in our experiment, more than 94\% transactions never diverge from their original gas consumption.
6\% of transactions are deviating, a large fraction ($>2.8\%$) deviates in 40\%-60\% of their executions and three out of four deviating transactions are privately mined.
The data are coherent when arguing that a block with 168 transactions on average has 4 transactions with at least one deviating execution on average.
Interestingly, according to our data, blocks contain on average 10 deviating transactions.

We conjecture that most transactions have a constant gas consumption because Ethereum provides no inclusion or ordering guarantees for transactions and transaction senders typically only sign transactions that will succeed regardless of block inclusion and ordering.
Hence, most transactions are not ordering dependent and consume the same amount of gas in every permuted execution.


\subsubsection{Execution-focused Results}
\label{overallBlockPermutations}
In this section, we evaluate data per execution:
We calculate metrics for each execution \textbf{without} grouping them according to blocks or transactions.
Consequently, the visualizations plot the metrics over all recorded executions.

\paragraph{Intensity of Deviation}
Our second metric $i^\mathrm{R}_E$ measures the intensity of the gas deviation of block executions $E_\mathrm{B}$ and transaction executions $E_\mathrm{T}$ relative to their original gas consumption $g$ according to expressions~\ref{eqRelPermBlockGasDiff} and~\ref{eqRelPermTxsGasDiff}.

\begin{minipage}{0.45\linewidth}
\begin{equation}
\label{eqRelPermBlockGasDiff}
    i^\mathrm{R}_{E_\mathrm{B}} = \frac{g_{E_\mathrm{B}}}{g_\mathrm{B}}
\end{equation}
\end{minipage}
\begin{minipage}{0.45\linewidth}
\begin{equation}
\label{eqRelPermTxsGasDiff}
    i^\mathrm{R}_{E_\mathrm{T}} = \frac{g_{E_\mathrm{T}}}{g_\mathrm{T}}
\end{equation}
\end{minipage}

Values for $i^\mathrm{R}_{E_\mathrm{B}}$ must be in the interval $[0.12\%, 85714.29\%]$ and values for $i^\mathrm{R}_{E_\mathrm{T}}$ must be in the interval $[0.06\%, 171,428.57\%]$.\footnote{Based on the minimum of 21,000 units of gas for transactions and the new block gas limit of 36,000,000 for blocks which we use throughout the paper.}

\begin{figure}[htbp]
    \centering
    \begin{tikzpicture}
\begin{axis}[
    height=5cm,
    width=8.5cm, 
    log origin=infty,
    ytick={0, 0.0001, 0.001, 0.01, 0.1, 1.0, 10.0, 100.0},   
    ylabel={Share in block permutations},                  
    ylabel style={yshift=8pt},       
    xlabel={Relative gas consumption intervals},                 
    xlabel style={yshift=-50pt, xshift=-5pt},       
    xtick={0, 1, 2, 3, 4, 5, 6, 7, 8, 9, 10, 11, 12, 13, 14, 15, 16, 17, 18, 19, 20, 21},              
    xticklabels={
    \text{[0\%, 10\%)},
    \text{[10\%, 20\%)},
    \text{[20\%, 30\%)},
    \text{[30\%, 40\%)},
    \text{[40\%, 50\%)},
    \text{[50\%, 60\%)},
    \text{[60\%, 70\%)},
    \text{[70\%, 80\%)},
    \text{[80\%, 90\%)}, 
    \text{[90\%, 100\%)}, 
    \text{100\% (same)}, 
    \text{(100\%, 110\%]},
    \text{(110\%, 120\%]},
    \text{(120\%, 130\%]},
    \text{(130\%, 140\%]},
    \text{(140\%, 150\%]},
    \text{(150\%, 160\%]},
    \text{(160\%, 170\%]},
    \text{(170\%, 180\%]},
    \text{(180\%, 190\%]}, 
    \text{(190\%, 200\%]},
    \text{(200\%, $\infty$\%]}
    }, 
    yticklabels={0.0001\%, 0.001\%, 0.01\%, 0.1\%, 1\%, 10\%, 100\%}, 
    ymin=0, ymax=100,                  
    bar width=6pt,                    
    enlarge x limits={abs=0.75cm},     
    ymajorgrids=true,                  
    grid style=dashed,                  
    xticklabel style={rotate=90},      
    ybar,                               
    ymode=log,
    nodes near coords,                 
    nodes near coords align={vertical},
    every node near coord/.append style={font=\tiny, rotate=90, xshift=17px, yshift=-5px}, 
    point meta=rawy
]
\addplot coordinates {
    (0, 0.000) 
    (1, 0.000024) 
    (2, 0.0002) 
    (3, 0.001) 
    (4, 0.0035) 
    (5, 0.0163) 
    (6, 0.0696) 
    (7, 0.2189) 
    (8, 0.9587) 
    (12, 0.1878) 
    (13, 0.0209) 
    (14, 0.0076) 
    (15, 0.0028) 
    (16, 0.0021) 
    (17, 0.0015) 
    (18, 0.0022) 
    (19, 0.0013) 
    (20, 0.0002) 
    (21, 0.0046) 
};
\addplot+[color=Bittersweet,fill=Dandelion,forget plot] coordinates {
    (8, 57.6806) 
    (9, 21.3834) 
    (10, 19.4366) 
};
\end{axis}
\end{tikzpicture}
    \caption{Distribution of block permutations according to the gas deviation intensity $i^\mathrm{R}_{E_\mathrm{B}}$.
    Note the log-scaled y-axis and x-axis interval ranges.
    Orange-colored bars represent more than 98\% together. 
    }
    \label{figRelativeBlockGasDistribution}
\end{figure}
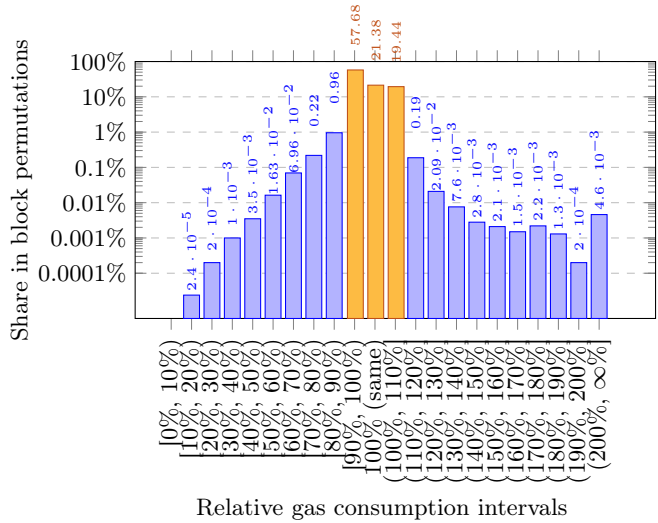

Fig.~\ref{figRelativeBlockGasDistribution} shows the empirical distribution of $i^\mathrm{R}_{E_\mathrm{B}}$ over all 262 million block permutations in the database.
Values are in the range $[19.78\%, 363.88\%]$ with an average of $99.23\%$ and standard deviation of $2.88\%$.
With more than 98\%, almost all block permutations deviate at most 10\% from their original gas consumption.
Note that invalid executions (from block permutations with protocol violations) are not included because we do not have their actual gas consumption.


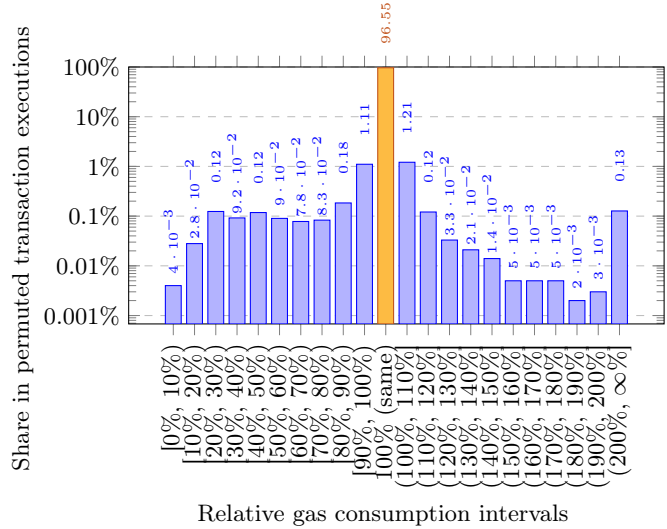
\begin{figure}[htbp]
    \centering
    \begin{tikzpicture}
\begin{axis}[
    height=5cm,
    width=8.7cm,
    log origin=infty,
    ytick={0, 0.0001, 0.001, 0.01, 0.1, 1.0, 10.0, 100.0}, 
    ylabel={Share in permuted transaction executions},            
    ylabel style={yshift=5pt,xshift=-20pt},      
    xlabel={Relative gas consumption intervals},                
    xlabel style={yshift=-50pt, xshift=-5pt},       
    xtick={0, 1, 2, 3, 4, 5, 6, 7, 8, 9, 10, 11, 12, 13, 14, 15, 16, 17, 18, 19, 20, 21},             
    xticklabels={
    \text{[0\%, 10\%)},
    \text{[10\%, 20\%)},
    \text{[20\%, 30\%)},
    \text{[30\%, 40\%)},
    \text{[40\%, 50\%)},
    \text{[50\%, 60\%)},
    \text{[60\%, 70\%)},
    \text{[70\%, 80\%)},
    \text{[80\%, 90\%)}, 
    \text{[90\%, 100\%)}, 
    \text{100\% (same)}, 
    \text{(100\%, 110\%]},
    \text{(110\%, 120\%]},
    \text{(120\%, 130\%]},
    \text{(130\%, 140\%]},
    \text{(140\%, 150\%]},
    \text{(150\%, 160\%]},
    \text{(160\%, 170\%]},
    \text{(170\%, 180\%]},
    \text{(180\%, 190\%]}, 
    \text{(190\%, 200\%]},
    \text{(200\%, $\infty$\%]}
    }, 
    yticklabels={0.0001\%, 0.001\%, 0.01\%, 0.1\%, 1\%, 10\%, 100\%}, 
    ymin=0, ymax=100,                 
    bar width=6pt,
    ymajorgrids=true,          
    grid style=dashed,               
    xticklabel style={rotate=90},     
    ybar,                               
    ymode=log,
    nodes near coords,                 
    nodes near coords align={vertical},
    every node near coord/.append style={font=\tiny, rotate=90, xshift=17px, yshift=-5px}, 
    point meta=rawy
]
\addplot coordinates {
    (0, 0.004) 
    (1, 0.028) 
    (2, 0.124) 
    (3, 0.092) 
    (4, 0.118) 
    (5, 0.09) 
    (6, 0.078) 
    (7, 0.083) 
    (8, 0.184) 
    (9, 1.105) 
    (11, 1.212) 
    (12, 0.121) 
    (13, 0.033) 
    (14, 0.021) 
    (15, 0.014) 
    (16, 0.005) 
    (17, 0.005) 
    (18, 0.005) 
    (19, 0.002) 
    (20, 0.003) 
    (21, 0.127) 
};
\addplot+[color=Bittersweet,fill=Dandelion,forget plot] coordinates {
    (9, 96.547) 
};
\end{axis}
\end{tikzpicture}
    \caption{Distribution of permuted transaction executions according to the gas deviation intensity $i^\mathrm{R}_{E_\mathrm{T}}$.
    Note the log-scaled y-axis and x-axis interval ranges.
    The orange-colored bar represents more than 96\%.
    }
    \label{figRelativeTransactionGasDistribution}
\end{figure}

Fig.~\ref{figRelativeTransactionGasDistribution} shows the distribution of $i^\mathrm{R}_{E_\mathrm{T}}$ over all 42 billion permuted transaction executions in the database.
Values are in the range $[0.31\%, 31,856.31\%]$ with an average of $100.04\%$ and standard deviation of $14.66\%$.
With 96\%, most permuted transaction executions do not deviate from their original gas consumption.

According to these results, only very few block permutations result in significant deviations in gas consumption.
However, even less intense gas deviations can result in gas limit reached errors and hence invalid executions as shown in Sec.~\ref{protocolViolations}. 
In addition, the previously stated conjecture is also supported by the results.

\section{Discussion and Future Work}
\label{discussion}
The data we used for the experiment (i.e. Ethereum Mainnet blocks, transactions and states) reflect the current usage of Ethereum.
The results of the experiment are only valid for the observed usage which can change in the future for various reasons.
Also, we executed at most 1000 permutations per block because computing all permutations is infeasible for most blocks.
In addition to computational complexity, functionality to execute custom blocks is challenging too.
Common client software is shipped without the capability to consecutively execute multiple transactions on a specific state so that incremental state changes are taken into account.
Available developer tools provide functionality to trace the execution of only a single transaction on a specific state with custom overrides and to simulate a block on a given state.
The issue with the latter is the missing ability to export a specific state from the Mainnet without huge storage requirements.
However, the community realized there is a need to simulate custom blocks on specific states and is actively working on a standardized functionality to simulate even multiple blocks with the `simulateV1' API~\cite{Killari2023Eth_multicallV1,KillariDev2023Add484}.

The experiment could be enhanced by capturing, augmenting, and analyzing execution traces (sequences of the performed steps during the state transition) to develop a deep understanding of the reasons for gas deviations and protocol violations in block permutations by identify resposible applications and usage patterns.
We used Tenderly\footnote{https://tenderly.co/transaction-simulator {\scriptsize (2025-05-20)}} to inspect execution traces of two  transactions for which we recorded the most significant gas deviation.
The first transactions consumed just 0.3\% (44k gas) of its original gas consumption (14M gas), in which the transaction executed a set of methods of the UniswapV3 application many times until reaching its gas limit.
The second transaction represents the other end of the scale and through the traces, we see that the same contracts as with the first transaction are involved but the transaction consumed 318 times (14M gas) of its original gas consumption (44k gas) in which the transaction reverted.
An augmentation of execution traces, however, requires consideration of ethical aspects that go beyond the experiment described in this paper.

Another way of using execution traces is to determine conflicts between transactions~\cite{Biton2025,Anjana2025Empirical,Sei2024}.
The purpose of such analysis is to quantify the potential for parallel transaction execution and to find a scheduling well suited for parallel execution.
Our experiment shows that the resource consumptions in terms of gas varies depending on the scheduling and must be considered by scheduling algorithms to achieve the best performance.
We consider this line of research as future work, which should also incorporate methods from program analysis to better capture the dynamic execution behavior of transactions.

Introducing randomized transaction orderings in practice requires the verifiable and unpredictable generation of a seed value to permute transactions, an approach is presented in~\cite{Droll2024ShortEthereum}.
Also, based on our results, further aspects need to be handled appropriately.
Reaching the block gas limit, as in 6\% of block permutations in the experiment, might be acceptable because block assemblers can manage the risk of block invalidation by putting less transactions in a block, as our data shows.
To avoid invalid blocks due to insufficient funds, the block assembler can include at most one transaction per sender, who must have sufficient funds.
However, the recently introduced account abstraction~\cite{Buterin2024EIP-7702} requires modifications to this method to guarantee that sending accounts do not lose funds due to other transactions in the block.
Economic effects on stakeholders should also be evaluated.

\section{Conclusion}
\label{conclusion}
In this empirical study, we permuted transactions of Ethereum blocks to quantify the effects of intra-block dependencies regarding protocol violations, execution errors and gas consumption. 
Invalid block permutations due to protocol violations can be attributed to exceeded gas limits and insufficient funds to settle fees.
Permuting transactions does not affect the observed execution error types but they occur slightly more often than in original blocks.
The gas consumption of most transactions remains constant under permutation, and the gas consumption of blocks is mostly stable as long as the block limit is not exceeded.
Privately mined transactions stand out in all analyzed categories.
They lead to insufficient fund errors, account for new failing transactions and are responsible for a large share of gas deviating transactions.
Based on these findings, we suppose that senders of privately mined transactions trust on a specific order or position of their transactions in blocks while senders of other transactions predominantly do not.
Our findings support the technical practicability of randomized transaction ordering.
Block producers can manage to not exceed the block gas limit with gas margins and by avoiding privately mining transactions, and eliminate insufficient funds with a content-aware transaction selection strategy.
Nonce-related errors are prevented with the present permutation algorithm. 
However, avoiding privately mined transactions would constitute a strong and disruptive step — one likely to face resistance and require further justification.

Future work might identify applications and usage patterns responsible for gas deviations since they are also relevant for parallel execution.

%
%

\bibliographystyle{splncs04}
\bibliography{references-manually}

\end{document}